\documentclass[a4paper,twoside]{article}
\newcommand{\affil}[1]{$^{\rm #1}$}
\usepackage{amssymb}
\usepackage[authoryear]{natbib}
\usepackage{url}
\bibpunct{(}{)}{;}{a}{}{,}
\usepackage{graphicx}
\date{} 
\title{\large\bf\flushleft The Parkes Observatory Pulsar Data Archive}
\author{\parbox{\textwidth}{\flushleft
\vspace{-0.5cm}
{\it G. Hobbs\affil{A}, D. Miller\affil{B}, R. N. Manchester\affil{A}, J. Dempsey\affil{B}, J. M. Chapman\affil{A}, J. Khoo\affil{A}, J. Applegate\affil{B}, M. Bailes\affil{C},  N. D. R. Bhat\affil{C}, R. Bridle\affil{B}, A. Borg\affil{B}, A. Brown\affil{A}, C. Burnett\affil{D},  F. Camilo\affil{E}, C. Cattalini\affil{B}, A. Chaudhary\affil{A}, R. Chen\affil{B}, N. D'Amico\affil{F}, L. Kedziora-Chudczer\affil{G}, T. Cornwell\affil{A}, R. George\affil{B}, G. Hampson\affil{A}, M. Hepburn\affil{B}, A. Jameson\affil{C}, M. Keith\affil{A}, T. Kelly\affil{B}, A. Kosmynin\affil{A}, E. Lenc\affil{A}, D. Lorimer\affil{H}, C. Love\affil{B}, A. Lyne\affil{I}, V. McIntyre\affil{A}, J. Morrissey\affil{B}, M. Pienaar\affil{B}, J. Reynolds\affil{A}, G. Ryder\affil{B}, J. Sarkissian\affil{A}, A. Stevenson\affil{B}, A. Treloar\affil{J}, W. van Straten\affil{C}, M. Whiting\affil{A}, G. Wilson\affil{B}}\\
\vspace{0.4cm}
{\small \affil{A}\, CSIRO Astronomy and Space Science, P.O. Box 76, Epping, NSW 1710 Australia.}\\
{\small \affil{B}\,  CSIRO Information Management \& Technology (IM\&T), PO Box 225, Dickson ACT 2602.}\\
{\small \affil{C}\, Centre for Astrophysics and Supercomputing, Swinburne University of Technology, P.O. Box 218, Hawthorn VIC 3122, Australia.}\\
{\small \affil{D}\, University of Melbourne, VIC, Australia.}\\
{\small \affil{E}\, Columbia Astrophysics Laboratory, Columbia University, New York, NY 10027, USA.}\\
{\small \affil{F}\, INAF - Osservatorio Astronomico di Cagliari, Poggio dei Pini, 09012 Capoterra, Italy.}\\
{\small \affil{G}\, School of Physics, UNSW, Sydney, NSW 2052 Australia.}\\
{\small \affil{H}\, Department of Physics, West Virginia University, Morgantown, WV 26506, USA.}\\
{\small \affil{I}\, Jodrell Bank Centre for Astrophysics, University of Manchester, Manchester, M13 9PL, UK.}\\
{\small \affil{J}\, Australia National Data Service, Monash University, 680 Blackburn Road, Clayton, VIC 3168, Australia.}\\
}
}
\begin{document}
\twocolumn[
\begin{changemargin}{.8cm}{.5cm}
\begin{minipage}{.9\textwidth}
\vspace{-1cm}
\maketitle
%
%
\small{\bf Abstract:}
The Parkes pulsar data archive currently provides access to 144044 data files obtained from observations carried out at the Parkes observatory since the year 1991.  Around 10$^5$ files are from surveys of the sky, the remainder are observations of 775 individual pulsars and their corresponding calibration signals.  Survey observations are included from the Parkes 70\,cm and the Swinburne Intermediate Latitude surveys.  Individual pulsar observations are included from young pulsar timing projects, the Parkes Pulsar Timing Array and from the PULSE@Parkes outreach program. The data files and access methods are compatible with Virtual Observatory protocols. This paper describes the data currently stored in the archive and presents ways in which these data can be searched and downloaded.  

\medskip
\medskip
\end{minipage}
\end{changemargin}
]
\small

\section{Introduction}

Observations of pulsars have provided insight into many areas of physics and astronomy. Such observations allowed the discovery of extra-Solar planets (Wolszczan \& Frail 1992)\nocite{wf92},  provided evidence of gravitational wave emission (Taylor \& Weisberg 1982) and have been used to test the general theory of relativity  (Kramer et al. 2006)\nocite{ksm+06}.   Pulsars are still being discovered (e.g., Keith et al. 2010).  These, and previously known pulsars,  are observed for many research projects with aims as diverse as detecting gravitational wave signals (e.g., Hobbs et al. 2010)\nocite{haa+10}, measuring the masses of objects in our Solar System (Champion et al. 2010), studying the interstellar medium (e.g., Hill et al. 2003, You et al. 2007) and determining the properties of the pulsars themselves (e.g., Lyne et al. 2010).

\begin{table*}
\caption{Receiver systems used for data in the archive.}\label{tb:receivers}
\begin{tabular}{lllllllll}
\hline
Name & Labels & $\nu$ & $\Delta \nu$ & N   & Data span & N$_f$  \\
 & & (MHz) & (MHz) & &   & \\
\hline
20cm multibeam & MULTI, MULT\_1$^a$ & 1369 & 288$^b$ & 13 & 09/2004--10/2010 & 87237  \\
H-OH & H-OH & 1405 & 256$^b$ & 1  & 02/2004--05/2007 & 2181 \\
1050cm$^b$ & 1050CM, 10CM & 3100 & 1024 & 1  & 02/2004--10/2010 & 5302  \\
1050cm$^b$ & 1050CM, 50CM  & 732 & 64 & 1 & 02/2004--10/2010 & 4112  \\
70cm & 70CM  & 430 & 32 & 1 & 05/1991--12/1994 & 42801  \\
\hline
\end{tabular}

$^a$ For some early files the MULT\_1 label is used to represent the central beam of the multibeam receiver \\
$^b$ Recent digital-filterbank systems provide 256\,MHz of bandwidth. However, both the multibeam receiver and the H-OH receiver can provide wider bandwidths. \\ 
$^c$ The 1050CM receiver is a dual-band receiver; see text. 
\end{table*}

Many pulsar observations have been obtained using National Facility telescopes which have little restriction on who may apply to carry out observations.  Time on such telescopes is usually awarded on the basis of the scientific merit of an observing proposal. Policies exist at most of these telescopes to make the resulting data available for the general scientific community after a specified period.  However, because of the amount of data, the complexity of the data formats, lack of storage space and because pulsar astronomers often develop their own hardware for data acquisition, it is difficult for non-team members to obtain such data sets after the embargo period. 

Numerous new scientific results have resulted from re-processing historical data. For instance, a re-analysis of a pulsar survey in the Magallenic Clouds led to the discovery of a single burst of radio emission that may be extra-Galactic in origin (Lorimer et al. 2007)\nocite{lbm+07}.  The Parkes multibeam pulsar survey (Manchester et al. 2001) has been re-processed numerous times which, to date, has led to the discovery of a further $\sim$30 pulsars (Eatough et al. 2010, Keith et al. 2009) and 10 new rotating radio transients (Keane et al. 2010).

In order to simplify access to astronomical data sets the ``Virtual Observatory'' (VO) was created\footnote{\url{http://www.ivoa.net/}}.  The VO aims to provide protocols for the storage, transfer and access of astronomical data and is commonly used for astronomical catalogues, images and spectral data.    The standard data formats used by the VO are the VOTable\footnote{\url{http://www.ivoa.net/Documents/VOTable/}} and the Flexible Image Transport System (FITS; Hanisch et al. 2001).  Hotan, van Straten \& Manchester (2004)\nocite{hvm04} extended FITS to provide a data storage structure that is applicable for pulsar data (this format is known as PSRFITS). The PSRFITS format allows pulsar observations to be analysed using VO tools. However, to date, the pulsar community has not extensively used such tools.

We have developed a data archive that will contain most of the recoverable pulsar observations made at the Parkes Observatory. The data (both the metadata describing the observations and the recorded signal from the telescope) have all been recorded in, or converted to, a common standard and the entire archive system has VO capabilities.  In this paper we first describe the observing systems at the Parkes observatory (\S2), the data formats used and the observations currently available from the data archive (\S3), tools available for searching and accessing the data (\S4), software that may be used with the data sets (\S5) and a description of the anticipated longer-term development of the data archive (\S6).  

\begin{table*}
\caption{Backend instrumentation at Parkes for which data are included in the archive.}\label{tb:bkends}
\begin{tabular}{p{4cm}lllll}
\hline
Name & Label & Maximum & Mode &Data span & N$_f$\\
 & & Bandwidth (MHz) & & & \\
\hline
Filterbank & AFB\_32\_256 & 32 & S & 05/1991--12/1994 & 42801\\
Analogue filterbank & AFB & 576 & S & 02/1998--05/2007 & 75304 \\
Wide band correlator & WBCORR & 256 & F & 02/2004--03/2007 & 2858 \\
Parkes digital filterbank 1 & PDFB1 & 256 & F & 12/2005--01/2008 & 2388 \\
Parkes digital filterbank 2 & PDFB2 &  256 & F & 04/2007--05/2010 & 2249 \\
Parkes digital filterbank 3 & PDFB3 &  1024 & FS & 03/2008--10/2010 & 2886 \\
Parkes digital filterbank 4 & PDFB4 &  1024 & FS & 10/2008--10/2010 & 2201 \\
Caltech-Parkes-Swinburne-Recorder 2 & CPSR2 & 2$\times$64 & F & 12/2002--06/2010 & 13354\\
\hline
\end{tabular}
\end{table*}

\section{Observing systems}

All data currently available from the archive were obtained using the Parkes 64-m radio telescope. The observing system used for pulsar observations is typically divided into the ``frontend" system, which includes the receiver and the ``backend'' system which refers to the hardware used to record and process the signal. 

Even though the Parkes telescope allows for multiple receivers to be installed on the telescope simultaneously, only one frontend can be used for a given observation.  In order to increase the survey speed of the telescope various multibeam receivers have been developed.  For instance, the 20\,cm multibeam receiver  (Staveley-Smith et al. 1996) allows 13 independent patches of the sky to be observed simultaneously (referred to as 13 ``beams'').   The changing lines of sight to radio pulsars leads to dispersive delays that are time-dependent. To remove these delays, simultaneous observations at two widely-spaced frequencies are desirable.   A dual-band receiver has been developed that allows simultaneous observations in the 10\,cm and 50\,cm bands (Granet et al. 2005). A listing of the receiver systems that have been used for the pulsar observations included in the archive are given in Table~\ref{tb:receivers}. In column order, we provide the name of the receiver, a label describing the receiver, its current central frequency, the maximum bandwidth that the backend instrumentation processed, the number of available beams, the data span available and the number of files in the archive that made use of this receiver. Many of these receivers have been upgraded over time. For instance, it was necessary to modify the central observing frequency for the 50\,cm receiver from 685\,MHz to 732\,MHz because of digital television transmissions. 



In order to maximise the signal-to-noise ratio of any pulsar observation it is necessary to observe with wide bandwidths.  When processing such observations it is essential to remove the effect of interstellar dispersion.  This is often done by dividing the observing bandwidth into frequency channels.  However, each frequency channel is still affected by the interstellar dispersion.  It is possible to remove the dispersion entirely by recording the raw signal voltage and convolving with the inverse of the transfer function of the interstellar medium.  This is known as ``coherent dedispersion'' and, as this is computationally intensive, has only recently being applied to data with large (e.g., $\sim$256\,MHz) bandwidths.  

When searching for new pulsars (``search-mode'' observations),  the signal from the telescope is divided into multiple frequency channels, digitised and recorded at a specified sampling rate.  For most of the data sets currently in the archive, only one-bit samples are recorded and the two polarisation data streams simply summed to produce total intensity using an analogue filterbank system (Manchester et al. 2001). Several generations of an analogue filterbank system have existed at Parkes.  The first generation system is labelled ``AFB\_32\_256'' and provided a bandwidth of 32\,MHz and 256 frequency channels.  For later generations, the backend is simply labelled as the ``AFB''.  If a pulsar is discovered in a search-mode file then the same data can subsequently be ``folded" at the topocentric period of the pulsar in order to produce a single pulse profile for the pulsar.  

The average of many thousands of individual pulses produces an ``average pulse profile'' that is usually stable and is characteristic of the pulsar.   As the pulsar's period may not be known with sufficient precision (or the pulsar may be in a fast binary system) it is common to fold only short sections of the data (typically one-minute sections) as the data are recorded.  Subsequent processing can be undertaken to sum these ``integrations'' with a more accurate pulsar ephemeris.  The data archive contains ``folded'' observations from numerous observing systems.  The Caltech-Parkes\--Swin\-burne-Recorder (CPSR2; Bailes 2003; Hotan 2006) coherently de-dispersed the data and usually produced two data files each with 64\,MHz of bandwidth.  CPSR2 was decommissioned in June 2010 and replaced by the ATNF-Parkes-Swinburne-Recorder (APSR; van Straten \& Bailes 2010) which provides up to 1\,GHz of coherently de-dispersed data.  The archive also includes data from a wide-bandwidth correlator and the suite of Parkes digital filterbank systems (PDFB1, PDFB2, PDFB3 and PDFB4) (Manchester et al., in preparation).  Details of these instruments are listed in Table~\ref{tb:bkends} providing the name of the backend and its label, the maximum bandwidth that the backend can process, whether it is used in ``Search-mode'' (S) or ``Fold-mode'' (F), data span and the number of observations included in the archive. The PDFB systems record all data as PSRFITS files.   Data files from other instruments have been converted to PSRFITS before inclusion in the data archive.

\section{Data sets and data format}

\begin{table*}
\caption{Data currently stored in the archive.}\label{tb:dsets}
\begin{footnotesize}
\begin{tabular}{p{2.5cm}lllp{2.0cm}p{2.5cm}ll}
\hline
Project & ID  & N$_f$ & Status & Receiver & Backends & Median file size & Data span\\
 & & & & & & \\
\hline
70cm pulsar survey & P050  & 42801 & c & 70CM & AFB\_32\_256 & 18 MB & 05/1991--12/1994\\
Young pulsar timing & P262  & 4512 & c & MULTI, H-OH & AFB & 0.3 MB & 02/1998--05/2007  \\
Swinburne Intermediate latitude survey & P309 & 70792 & c&  MULTI & AFB & 25 MB& 06/1998--03/1999\\
Parkes Pulsar Timing Array & {\bf P456} & 25610$^a$ & o & MULTI, H-OH, 1050CM & WBCORR, PDFB1, PDFB2, PDFB3, PDFB4, CPSR2, APSR & 64$^a$ MB & 02/2004--10/2010\\
PULSE@Parkes & {\bf P595}  & 329$^a$ & o & MULTI & PDFB2, PDFB3, PDFB4& 56$^a$ MB& 04/2008--11/2010 \\
\hline
\end{tabular}

$^a$ Not including the calibration files
\end{footnotesize}
\end{table*}

Currently the archive contains data that have been recovered from five observing projects.  A summary of these data sets is given in Table~\ref{tb:dsets} and details are provided below.  In Table~\ref{tb:dsets} we provide the project name and reference (identifiers in bold represent continuing projects), N$_f$ the number of raw data files currently in the database, the status of the project ('o' for on-going projects and 'c' for completed projects), the receiver and backend instrumentation used, typical individual file sizes and the date of the first and last observation stored in the archive\footnote{Note that data have not always been recorded with the correct project identifier.   We recommend that, if possible, the project identification is confirmed with the observers before the data are referenced in a publication.}

All of the pulsar data stored in the data archive follow the PSRFITS standard (Hotan, van Straten \& Manchester 2004)\footnote{\url{http://www.atnf.csiro.au/research/pulsar/index.php?n=Main.Psrfits}}.   Each file contains a single observation of a pulsar or a particular area of sky; for observations using the 13-beam multibeam receiver, 13 separate PSRFITS files are produced for each telescope pointing.  We note that the PSRFITS definition allows the addition of new parameters when required and therefore older PSRFITS files may not include as much metadata as later files.  Prior to Version 2.10 the format was not fully compliant with Virtual Observatory standards.  We have therefore converted all such earlier files to the most up-to-date version of PSRFITS.  Even though a large number of parameters are stored in PSRFITS files many of these parameters are not useful as searchable metadata.    In Table~\ref{tb:metadata} we list the parameters that are recorded as part of the data archive and can be used in order to identify an observation of interest (for instance, searches can be carried out on the telescope position, but not on the attenuator settings for that observation).  Note that only the pulsar J2000 names are stored.  We provide no facility to search on the older B1950 names.  The ATNF Pulsar Catalogue (Manchester et al. 2005)\footnote{\url{http://www.atnf.csiro.au/research/pulsar/psrcat}} can be used to determine a pulsar's J2000 name.  

Each file was obtained as part of a specific observing programme that had been allocated observing time on a competitive basis. The relevant metadata describing the project was obtained from the original observing proposal requesting the use of the telescope.  We store the proposal abstract and names of researchers included on the proposal.  This was obtained and converted to ensure compliance with the VO protocols.

\begin{table*}
\caption{Searchable metadata stored for each file.}\label{tb:metadata}
\begin{footnotesize}
\begin{tabular}{ll}
\hline
Parameter label & Description \\
\hline
BACKEND & Backend instrument \\
BECONFIG & Backend configuration \\
DATE (CREATION-DATE)$^a$ & Date that the data file was created\\
DATE-OBS & Date of observation (YYYY-MM-DDThh:mm:ss UTC) \\
DEC (DEC\_ANGLE)$^a$ & Declination (dms). For the Virtual Observatory the angle is given in degrees. \\
FRONTEND & Name of the receiver \\
HDRVER & Version number for the PSRFITS format\\
MJD & Start time MJD \\
NRCVR & Number of receiver receptors \\
OBSBW & Bandwidth for observation (MHz) \\
OBSERVER & Initials for observer who carried out the observation\\
OBSFREQ & Central observing frequency (MHz) \\
OBSNCHAN & Number of frequency channels \\
OBS\_MODE & Pulsar, calibration or search \\
PROJID & Project identification code\\
RA  (RA\_ANGLE)$^a$ & Right ascension (hms). For the Virtual Observatory the angle is given in degrees. \\
SRC\_NAME & Source name or scan identifier \\ 
STT\_IMJD & Integer part of the MJD for the observation\\
STT\_LST & Start Local Sidereal Time (LST) \\
STT\_SMJD & Start time (sec. past UTC 00h) \\
STT\_OFFS & Offset in the start time (seconds)\\
TELESCOP & Telescope used for observation (currently all set to PARKES)\\
\hline
(FILENAME)$^a$ & Name of the data file \\
(FILESIZE)$^a$ & Size of the data file \\
(FILE\_LAST\_MODIFIED)$^a$ & Date and time for when the file was created or last modified\\
(OBSTYPE)$^a$ & Type of data file (raw, preprocessed or thumbnail image) \\
(OBS\_LENGTH)$^a$ & Total length of observation (in milliseconds) \\
\hline
\end{tabular}
\\
$^a$ If a different label is used within the PSRFITS file compared to Virtual Observatory searches then the Virtual Observatory label is given in parentheses.
\end{footnotesize}
\end{table*}

\subsection{Modification of the data files}

The data-archiving policy is that no further modifications are made to the raw data files after conversion to the PSRFITS format. In some cases new header parameters become available after the conversion to PSRFITS and such header metadata are updated, but the raw data are untouched. In rare cases it may become apparent that a mistake has been made in converting to PSRFITS from the raw tape or disk files.  In such cases the data files will be replaced with corrected versions.  The database stores information on when the last modification to any observation file has been made.

\subsection{Fold-mode observations}

\begin{figure}
\includegraphics[width=6cm,angle=-90]{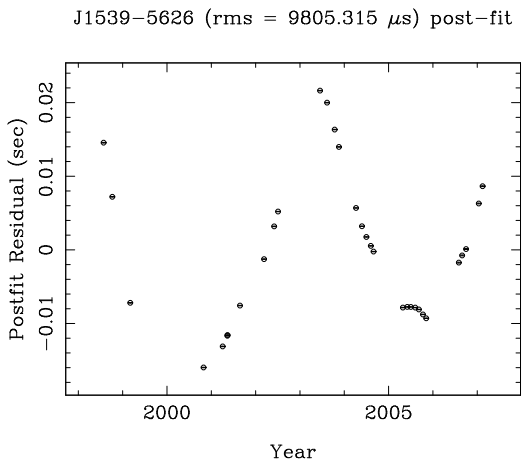}
\caption{Pulsar timing residuals for PSR J1539$-$5626 from the young pulsar timing programme, P262.}\label{fg:1539}
\end{figure}

\subsubsection{Young pulsar timing (Project code: P262)}

Long-term pulsar timing projects that have concentrated on pulsars with relatively small characteristic ages have been ongoing at Parkes for many years. Such projects have led to numerous publications on period glitches, pulsar timing irregularities and updated pulsar timing ephemerides (e.g., Wang et al. 2000).   Here we describe data from the P262 observing programme that was carried out between MJDs 50849 and 54224 (from Feb. 1998 to May 2007).  The data were recorded using the analogue filterbank system which records data in the search-mode format.  As these observations were of known pulsars the majority of the processing starts by folding the search-mode data at the known period of the pulsar\footnote{In a few cases it may be of interest to fold at a different period.  This could be because other pulsars were observed within the beam, to check whether the correct pulsar period is known or because the pulsar has ``glitched'' implying that the most recent ephemeris is not suitable for folding the data. The original search mode data will be made available, through this archive, at a later date and are currently available on request.}.  Data are available for 616 pulsars and were processed as follows:
\begin{itemize}
\item The original data files for all recoverable observations from the P262 observing programme were obtained.
\item The source name was updated to provide the most up-to-date name as presented in the ATNF Pulsar Catalogue.
\item The data were folded at the known period (using the most up-to-date pulsar ephemeris) of the pulsar using the DSPSR software (van Straten \& Bailes 2010) and a fold-mode PSRFITS file output.
\end{itemize}
In total, 4512 observations were recovered with a median observation time of five minutes and a total observation time of 597 hours. The observation filenames have a leading ``f" to indicate that they came from the analogue filterbank system followed by the date of the observation.  An example filename is ``f981007\_044636.rf" for an observation with a UTC start time of 1998 Oct 7, 04$^{\rm h}$46$^{\rm m}$36$^{\rm s}$.  As these data were obtained using the analogue fllterbank system we only provide total intensity profiles.

After the discovery of a pulsar, it is common to carry out a small number of ``gridding'' observations in order to improve the pulsar's position to a fraction of the telescope beamwidth (Morris et al. 2002).  For such observations the pulsar signal is often not observable, but such files can easily be identified as the telescope was not pointing directly at the pulsar.  

An example of the P262 data is shown in Figure~\ref{fg:1539}. This Figure contains the timing residuals (for details on the pulsar timing method see, e.g., Hobbs et al. 2006) obtained for a typical pulsar, PSR~J1539$-$5626.  For this pulsar 32 observations were observed as part of the P262 project over a period of 8.6\,yr.  The arrival time uncertainties are smaller than the symbol size in the figure and have a mean of $33\mu$s. The timing model used to determine the pre-fit timing residuals was obtained from the pulsar ephemeris stored in the PSRFITS file.  The data were first processed using the PSRCHIVE  (Hotan, van Straten \& Manchester 2004) software suite. First, the program \textsc{paz} was used to remove band edges and radio frequency interference (RFI) and \textsc{pam} was used to increase the signal-to-noise  ratio by integrating over the frequency channels and integrations). Pulse times-of-arrival were obtained using \textsc{pat} and finally timing residuals determined using \textsc{tempo2} (Hobbs, Edwards \& Manchester 2006).  The timing residuals are typical of normal pulsars that exhibit timing noise (cf., Hobbs et al. 2010).

\subsubsection{The PULSE@Parkes project (P595)}

\begin{table}
\caption{Pulsars observed as part of the PULSE@Parkes (P595) observing project.}\label{tb:p595}
\begin{footnotesize}
\begin{tabular}{llllll}
\hline
PSR J & Period & DM    & N$_f$ \\
             & (s)   & (cm$^{-3}$pc)\\ 
\hline
J0006$+$1834 & 0.694 & 12.0 & 6 \\
J0034$-$0721 & 0.943 & 11.38 & 18 \\
J0108$-$1431 & 0.807 & 2.38 & 14 \\
J0134$-$2937 & 0.137 & 21.81 & 10 \\
J0152$-$1637 & 0.833 & 11.92 & 10 \\ \\

J0206$-$4028 & 0.631 & 12.9 & 26 \\
J0437$-$4715 & 0.006 & 2.65   & 40 \\
J0452$-$1759 & 0.549 & 39.90 & 32 \\
J0729$-$1836 & 0.510 & 61.29 & 18 \\
J0742$-$2822 & 0.167 & 73.78 & 34 \\  \\

J0900$-$3144 & 0.011 & 75.70 & 16 \\
J0922$+$0638 & 0.431 & 27.27 & 10 \\
J0946$+$0951 &1.098 & 15.4  & 10 \\
J1003$-$4747 & 0.307 & 98.1 & 52 \\
J1107$-$5907 & 0.253 & 40.2 & 54 \\ \\

J1125$-$5825 & 0.003 & 124.78 & 6 \\
J1224$-$6407 & 0.216 & 97.47 & 32 \\
J1239$+$2453 & 1.382 & 9.24 & 6 \\
J1300$+$1240 & 0.006 & 10.17 & 6 \\
J1349$-$6130 & 0.259 & 284.6 & 26 \\ \\

J1359$-$6038 & 0.128 & 293.71 & 22 \\
J1412$-$6145 & 0.315 & 514.7 & 24 \\
J1453$-$6413 & 0.179 & 71.07 & 16 \\
J1530$-$5327 & 0.279 & 49.6 & 10 \\
J1543$-$0620 & 0.709 & 18.40 & 4 \\ \\

J1634$-$5107 & 0.507  & 372.8 & 12 \\
J1637$-$4553 & 0.119 & 129.23 & 8 \\
J1713$+$0747 & 0.005 & 15.99 & 2 \\
J1717$-$4054 & 0.888 & 307.09 & 17 \\
J1721$-$3532 & 0.280 & 496.0 & 12 \\ \\

J1726$-$3530 & 1.110 & 727.00 & 12 \\
J1807$-$2715 & 0.828 & 312.98 & 8 \\
J1818$-$1422 & 0.291 & 622.0 & 2 \\
J1829$-$1751 & 0.598 & 84.44 & 2 \\
J1820$-$0427 & 0.307 & 217.10 & 4 \\ \\

J1830$-$1059 & 0.405 & 161.50  & 2 \\
J1832$+$0029 & 0.534 & 28.3 & 6 \\
J1902$+$0615 & 0.674 & 502.90 & 4 \\
J2053$-$7200 & 0.341 & 17.3 & 6 \\
J2145$-$0750 & 0.016 & 9.00 & 10 \\ \\

J2317$+$1439 & 0.003 & 21.91 & 12 \\

\hline
\end{tabular}
\end{footnotesize}
\end{table}

The PULSE@Parkes project (Hobbs et al. 2009, Hollow et al. 2008) has been designed to introduce high school students to astronomy.  The students observe from a selection of $\sim$40 pulsars that are chosen to be of interest for various scientific projects.  The 20\,cm multibeam receiver is used, giving an observing frequency close to 1400\,MHz and a bandwidth of 256\,MHz.  Data have been recorded using the PDFB3 and PDFB4 backend systems.  Since the start of 2011, the PDFB3 system has been used to produce a high signal-to-noise pulse profile and simultaneously the PDFB4 system has recorded in search mode to provide information on single pulses and the RFI environment. Observations are typically 2 to 15\,min depending on the pulsar's flux density.  A pulsed  calibration signal is observed prior to each observation allowing each data set to be fully calibrated in polarisation and flux density. 

\begin{figure*}
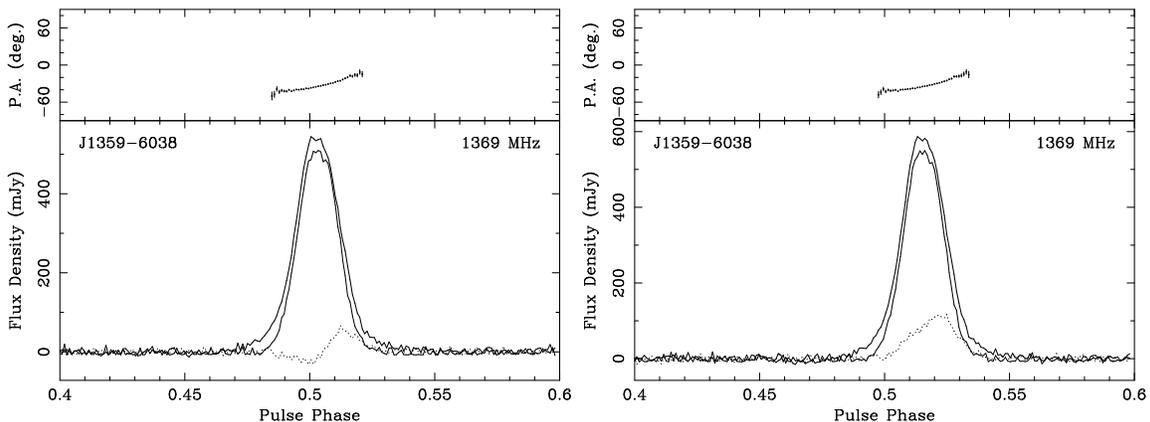

\includegraphics[width=5.5cm,angle=-90]{1359.ps}
\includegraphics[width=5.5cm,angle=-90]{1359_pcm.ps}
\caption{Profile for PSR J1359$-$6038 obtained by Kelso High School students as part of the PULSE@Parkes project. The profile in the left-hand panel has been calibrated using the standard \textsc{pac} calibration method.  The profile in the right-hand panel has been calibrated with compensation for cross-coupling in the 20\,cm feed.  The outer solid line represents Stokes I, the inner solid line the linear polarisation (with the position angle shown in the upper panel) and the dotted line shows Stokes V.}\label{fg:1359}
\end{figure*}

PULSE@Parkes  is an ongoing project and more data become available each month. As this project primarily has an outreach goal, these data sets are immediately available for download.   At the time of writing we have 661 observations from a total of 41 pulsars (listed in Table~\ref{tb:p595} which gives each pulsar's name, period, dispersion measure and the number of observations currently in the archive).  As for the P262 data, file names indicate the date and time of the observation.  File names starting with an ``r" correspond to PDFB2 data, ``s'' for PDFB3 data and ``t'' for PDFB4 data.  Folded pulsar archives have the file extension ``.rf''.  Calibration source files have the extension ``.cf'' and observations obtained in search mode have ``.sf''. In total 29\,GB of data are currently available for download.  We note that some of these pulsars are known to undergo extreme nulling events (during which the pulse disappears for many hours or days).   Some observations therefore seem to show no pulse.  Many of the other pulsars are affected by scintillation and, because of this, may have low signal-to-noise ratios in some observations.

\begin{figure}
\includegraphics[width=5.5cm,angle=-90]{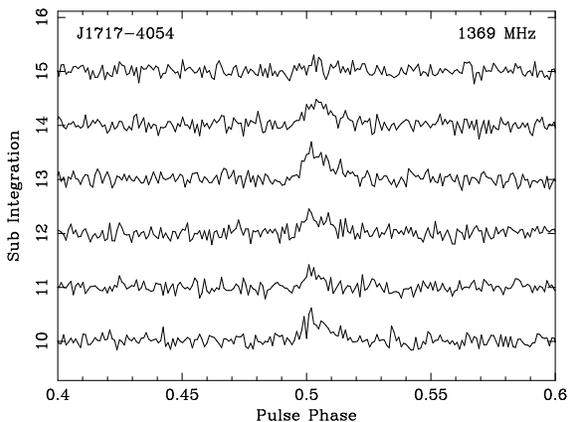}
\caption{Single pulses from the intermittent pulsar PSR~J1717$-$4054 obtained by students of the German International School Sydney as part of the PULSE@Parkes project.}\label{fg:1717}
\end{figure}

An example profile from the PULSE@Parkes project is shown in Figure~\ref{fg:1359}.  This pulse profile has been calibrated using \textsc{pac} in the PSRCHIVE software suite providing both polarisation and flux calibration.  An improved calibration method, described by van Straten (2004), uses feed cross-coupling data obtained using the program \textsc{pcm}.  The right panel in Figure~\ref{fg:1359} shows the pulse profile calibrated using the cross-coupling data, which agrees with that published by Karasteriou \& Johnston (2006). The differences between the two profiles in Figure~\ref{fg:1359} (particularly in Stokes V) highlight the importance of using careful calibration for observations obtained using the 20\,cm multibeam receiver.
An example of recent search mode PULSE@Parkes data are shown in Figure~\ref{fg:1717} where six adjacent individual pulses from the intermittent pulsar PSR~J1717$-$4054 are plotted. Many of the observations are affected by radio-frequency interference, but tools are available within the PSRCHIVE software suite to remove much of this interference.  

\subsubsection{The Parkes Pulsar Timing Array (P456)}

The Parkes Pulsar Timing Array (PPTA) project has the main aim of detecting gravitational wave signals (described in Verbiest et al. 2010,  Hobbs et al. 2009 and references therein). The main data collection for the project started in 2004 and is ongoing. Observations are taken every $\sim 3$\,weeks for 20 pulsars at three observing frequencies.  Several backend instruments are run in parallel.   This project makes extensive use of the 20\,cm multibeam receiver and the dual-band 10/50cm receiver.  Data have been recorded using an auto-correlation spectrometer (commonly referred to as the ``wide-bandwidth correlator'' and labelled as ``WBCORR''), coherent dedispersion systems (CPSR2 and APSR) and the digital filterbanks (PDFB1, PDFB2, PDFB3 and PDFB4).  Observations at the same time and frequency for different backends contain the same information and cannot be used as two independent observations of the pulsar. Data are recorded with a large number of frequency channels and typically one-minute integrations.  Polarisation information is available which can be calibrated to produce Stokes parameters.  Files have the same naming convention as in the P595 data with CPSR2 data at different frequencies denoted by an ``m'' or ``n'' at the start of the filename. 

The PDFB1/2/3/4 and WBCORR systems directly produce PSRFITS data and we make no changes to the data files for inclusion into the archive.  CPSR2 produces individual files for each integration for each observation.  We have combined these integrations into one PSRFITS file for each observation.  We have obtained the relevant metadata for the observation using (in most cases) the header information stored in simultaneous PDFB or WBCORR files.

\begin{figure*}
\begin{center}
\includegraphics[angle=-90,width=10cm]{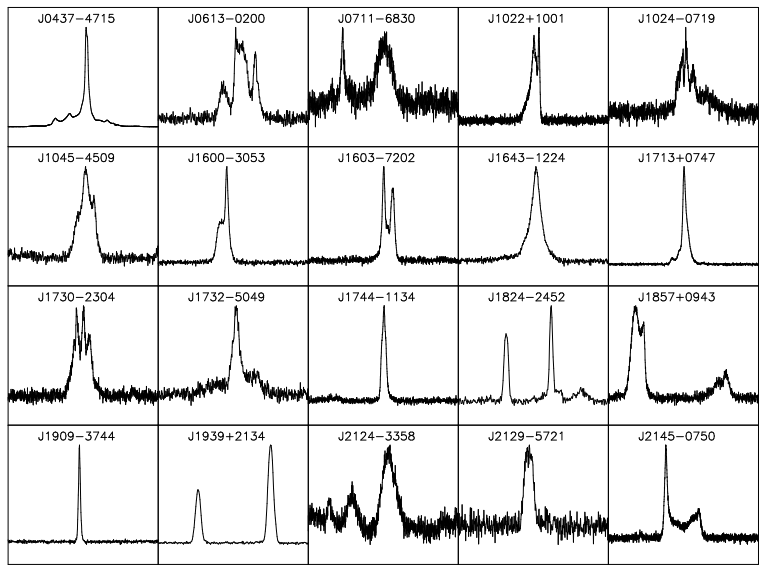}
\end{center}
\caption{Typical 20cm profiles from the PDFB4 backend for the Parkes Pulsar Timing Array pulsars obtained after a 1-hour observation.}\label{fg:p456prof}
\end{figure*}

Individual data files may be large.  Typical recent one-hour observations of PSR~J1022$+$1001 occupy 1.1\,GB.  The total amount of data provided as part of the archive is 3\,TB and this is expected to grow by $\sim$1\,TB/year.  
The period and dispersion measure of the pulsars observed as part of the project are given in Table~\ref{tb:p456} along with the total number of observations.   In Figure~\ref{fg:p456prof} we show typical total intensity pulse profiles in the 20\,cm observing band for each pulsar.

The data for this project can be used for numerous applications such as studying the polarisation properties of the pulsars (Yan et al. 2011), pulse shape variability or dispersion measure variations (You et al. 2007). However, getting the most from the data requires local knowledge of how the data were taken, issues with the backend systems during the observing, the local RFI environment, high quality standard templates etc.  This information is not provided as part of the data archive and we recommend that any users of these data sets obtain further information from the relevant PPTA papers (Verbiest et al. 2010,  Hobbs et al. 2009 and references therein).

\begin{table}
\caption{Pulsars observed as part of the Parkes Pulsar Timing Array (P456) observing project.}\label{tb:p456}
\begin{footnotesize}
\begin{tabular}{llllll}
\hline
PSR J & Period & DM    & N$_f$ \\
             & (ms)   & (cm$^{-3}$pc)\\ 
\hline
J0437$-$4715 & 5.757 & 2.65   & 4871 \\ 
J0613$-$0200 & 3.062 & 38.78 & 1372   \\
J0711$-$6830 & 5.491 & 18.41 & 1313   \\
J1022$+$1001 & 16.453 & 10.25 & 1539 \\
J1024$-$0719 & 5.162 & 6.49 & 1207  \\ \\
J1045$-$4509 & 7.474 & 58.15 & 1131\\
J1600$-$3053 & 3.598 & 52.19 & 1372 \\
J1603$-$7202 & 14.842 & 38.05 & 974 \\ 
J1643$-$1224 & 4.622 & 62.41 & 835 \\
J1713$+$0747 & 4.570 & 15.99 & 1054\\ \\
J1730$-$2304 & 8.123 & 9.61 & 752 \\
J1732$-$5049 & 5.313 & 56.84 & 597 \\
J1744$-$1134 & 4.075 & 3.14 & 1144 \\
J1824$-$2452 & 3.054 & 119.86 & 547 \\
J1857$+$0943 & 5.362 & 13.31& 668 \\ \\
J1909$-$3744 & 2.947 & 10.39 & 2167 \\
J1939$+$2134 & 1.558 & 71.04 & 677 \\
J2124$-$3358 & 4.931 & 4.62 & 1156 \\
J2129$-$5721 & 3.726 & 31.85 & 841 \\
J2145$-$0750 & 16.052 & 9.00 & 1115 \\
\hline
\end{tabular}
\end{footnotesize}
\end{table}

\subsection{Surveys}

\subsubsection{The 70cm pulsar survey (P050)}

The 70\,cm Southern-sky pulsar survey (Manchester et al. 1996, Lyne et al. 1998) led to the detection of 298 pulsars, of which 101 were new discoveries.  These discoveries included PSR~J0437$-$4715, the brightest millisecond pulsar known.   Each observation lasted 160\,s and 1-bit data were recorded with a sample interval of 300$\mu$s.  These survey observations were stored on $\sim 600$ exabyte tapes.  Some of these tapes are now unreadable, but, in total, we succeeded in recovering 42750 observations (93\% of the total survey). Each observation file is 18\,MB in size giving a total data storage of 935\,GB.  In addition to the survey observations, the tape files included 4263 re-pointings toward 293 different pulsars.   For each observation we have produced a single PSRFITS file.  We have included various parameters including the project code (P050), the label for the front-end receiver (70CM) and source name (either the pulsar name, or the pointing identifier) in the PSRFITS file.

In order to confirm that we have successfully converted the files to the PSRFITS format we have compared the results for a selection of observations bit-by-bit with the results obtained using the program, \textsc{sc\_td}, which was used during the original processing of the data.  No discrepancies were found. We have reprocessed all data using the search algorithm being used for the current Parkes HTRU pulsar survey (Keith et al. 2010).  All previously detected pulsars have been re-detected using the data stored in the archive.

\begin{figure}
\begin{center}
\includegraphics[width=8cm]{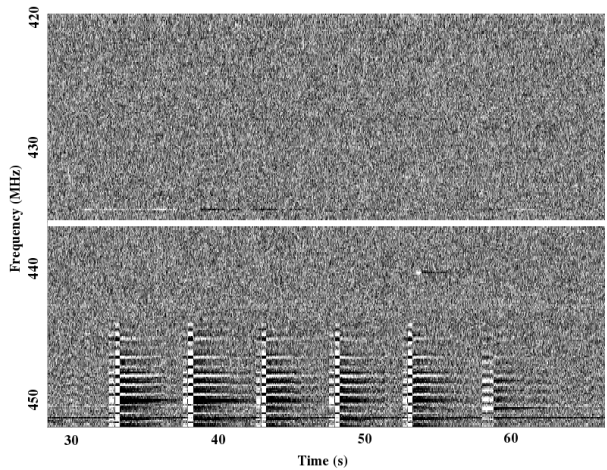}
\end{center}
\caption{Approximately 40\,seconds of data from the 70\,cm Parkes pulsar survey.  The high frequency channels in these data are affected by unexplained interference.}\label{fg:rfi}
\end{figure}

We note that all of the search mode data sets are in their original form and therefore contain imperfections, such as radio frequency interference.  For instance, we show in Figure~\ref{fg:rfi}, approximately 40 seconds of data for a typical observation.  The grey-scale image provides the intensity as a function of time and frequency.  It is clear that radio frequency interference is affecting the highest frequency channels (around a frequency of 450\,MHz).  Such interference needs to be identified and removed before standard search algorithms are applied to the data.

\subsubsection{The Swinburne Intermediate latitude survey (P309)}

These data are from a large survey for pulsars at high Galactic latitudes (Edwards, Bailes, van Straten \& Britton 2001).  The survey covered $\sim 4150$ square degrees in the region $-100^\circ \leq l \leq 50^\circ$ and $5^\circ \leq |b| \leq 15^\circ$ with 4702 pointings of the 13 beam receiver (providing 61126 individual files) each of 265\,sec. In total, 170 pulsars were detected of which 69 were new discoveries.  The raw data for this project are stored on Digital Linear Tape (DLT) at Swinburne University of Technology.  We were provided with data files for each observation that had been processed using the \textsc{sc\_td} software package. We converted each beam of each pointing to a single PSRFITS file and compared the converted files with the original files to ensure that the raw data was unchanged during the conversion process.  The PSRFITS header parameters were updated with the project code (P309), the telescope (PARKES), the receiver (MULTI) and the beam corresponding to the observation.

\begin{figure}
\begin{center}
\includegraphics[width=8cm]{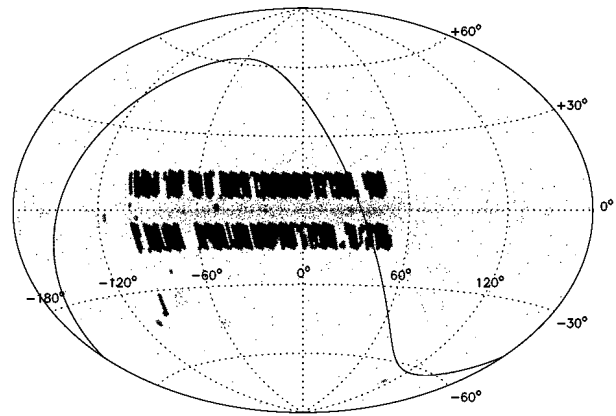}
\end{center}
\caption{Galactic coordinates for the Swinburne Intermediate Latitude Survey are indicated as bold points.  The area of the sky under the solid line is where the Parkes 70\,cm was conducted.  The small dots are the positions of known pulsars.}\label{fg:sky_coverage}
\end{figure}

This programme has 70792 observations stored in the archive.  These include most of the original survey observations and re-pointings toward detected pulsars. For survey observations the source name is set to ``Unknown'' and the pointing identification is set to a specific value unique to that particular observation.   In Figure~\ref{fg:sky_coverage} we plot the position of each observation that has been recovered overlaid on the positions of all known pulsars.

\section{Obtaining the data}

\begin{figure}
\begin{center}
\includegraphics[width=7cm]{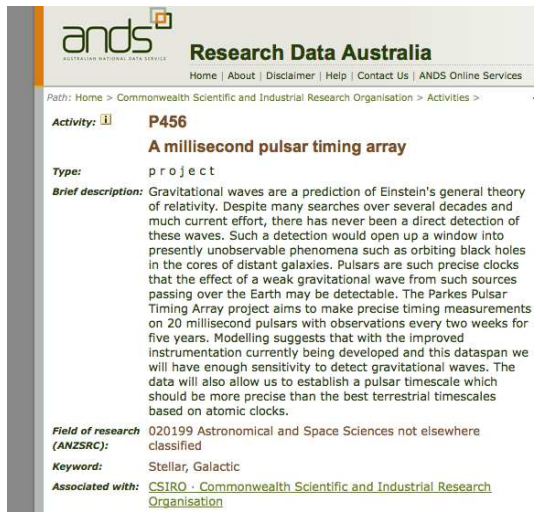}
\end{center}
\caption{Example screenshot from the ANDS portal that provides access to information about individual projects. }\label{fg:ands}
\end{figure}

\subsection{Data access portals}

The Parkes pulsar data archive can be accessed through various portals.  The Australia National Data Service (ANDS) portal, called Research Data Australia (RDA),\footnote{\url{http://www.ands.org.au; http://services.ands.org.au/home/orca/rda/.}} is used to search descriptions of data collections. CSIRO provides a data access portal\footnote{\url{http://datanet.csiro.au/dap/}} intended for use by professional astronomers to search for, and download, small numbers of data files. The PULSE@Parkes portal\footnote{\url{http://outreach.atnf.csiro.au/education/pulseatparkes/}} makes the data accessible to the broader community. Virtual Observatory tools can also be used to query the database.

\subsubsection{Research Data Australia portal}

The Australia National Data Service (ANDS) intends to present information about, and access to, as much Australian research data as possible in a common manner.  This portal can be used in order to obtain information about various pulsar projects and data collections. For instance, a user can search for ``astronomical data" and then obtain information on e.g., the P456 Parkes Pulsar Timing Array project.    Note that this portal will not  allow queries based on observational parameters such as the source name or position.  The emphasis of Research Data Australia (RDA) is on discovering the existence of collections of data, with discipline-specific queries being handled by specific portals such as those described below. An example is shown in Figure~\ref{fg:ands} where information is provided on the P456 project.  Note that the CSIRO Data Access Portal (described in \S\ref{sec:psrportal}) provides links to the relevant parts of the Research Data Australia website.

\subsubsection{The CSIRO Data Access Portal}\label{sec:psrportal}

\begin{figure*}
\begin{center}
\includegraphics[width=13cm]{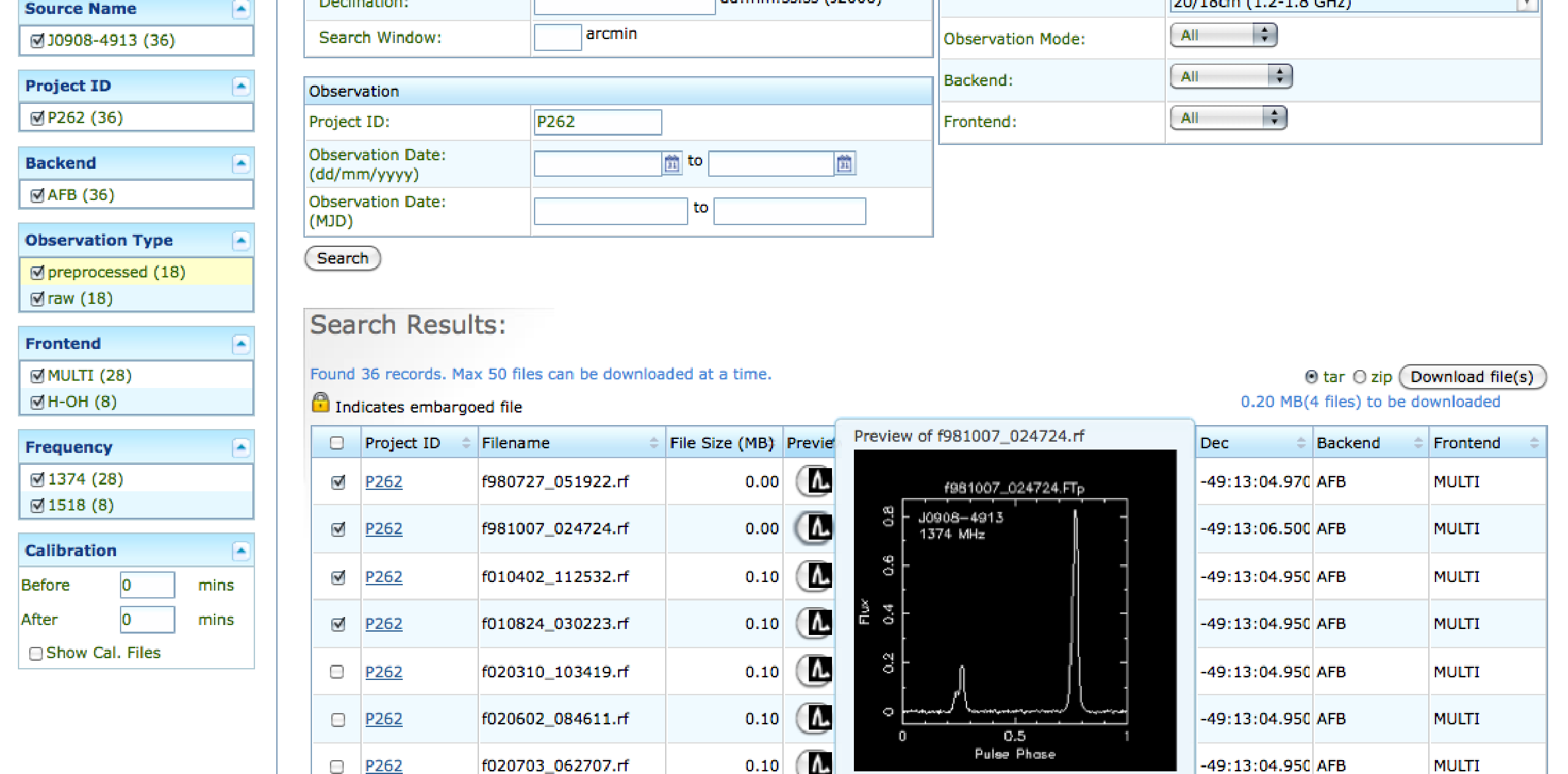}
\end{center}
\caption{Example screenshot from the CSIRO data access (pulsar) portal. The top panel allows the user to select sky-positions, a pulsar name, project identifier or date range to restrict the search results.  The panel on the left divides the search results into various subsections.  The bottom panel shows the result from a search and the thumbnail image gives an indication of the data quality.}\label{fg:csiro}
\end{figure*}

The CSIRO data access portal provides an interface to data sets including the Parkes pulsar observations.  This system allows searching on pulsar name, project identification or areas of the sky. An example screen-shot is shown in Figure~\ref{fg:csiro}. This portal provides a means to download a small number of individual files from the archive.  Typical usage would be to search for a particular pulsar name (e.g., ``J0437-4715").  At the time of writing, this returns 10008 files stored in the database.   These are divided into the original data files (5112 files) and pre-processed files (4896 files).   A panel is presented providing a basic description of  these files (e.g,.\,1072 observations were obtained using the PDFB1 system and 40 of these observations were obtained as part of the PULSE@Parkes project).  The user can then filter these results to obtain, for instance, only PULSE@Parkes observations, obtained with the PDFB4 backend instrumentation.  This reduces the number of files to 10 which can be selected for download.  

Most of the fold-mode observations have corresponding pre-processed files that have been summed in polarisation, frequency and time.  These pre-processed files are significantly smaller than the raw observations and can be used for many purposes.  However, it will not be possible to undertake any high-precision pulsar timing, frequency-dependent investigations nor analysis of the pulse polarisation using such data.   Thumbnail images of these pre-processed files are available.  These should be viewed before a file is selected for download to ensure that the data quality is sufficient for the project being undertaken.

If required, calibration files can also be downloaded.  As calibration files may have been obtained before or after the pulsar observation, the CSIRO data access portal provides the ability to download all calibration files within a specified time range before or after the start of the pulsar observation.

With a few exceptions, observations from the Parkes radio telescope are embargoed for a period of 18 months from the time that the data were obtained.   The CSIRO access portal is the only generally accessible means by which files can currently be downloaded and therefore requires the user to provide a user name and password if embargoed data are required.  An individual who is part of an observing project can log on to the portal using the account that they used to submit or view their observing proposal.  

\subsubsection{The PULSE@Parkes portal}

Simplified versions of the PULSE@Parkes data sets are also available from the project website.  This website provides images of each observation and the data in a simple text form that can loaded into a spreadsheet.  A simple web interface allows the data to be processed online to determine the pulsar dispersion measures and characteristic ages.  New online educational modules using these data sets will become available in the future.

\subsubsection{The Virtual Observatory Interface}

The Virtual Observatory (VO) allows a user to combine and compare a large number of different data sets.  A diverse range of astronomical catalogues and images are already available through the VO including pulsar catalogues and the tables of pulsar parameters that have been included in recent publications.  The International Virtual Observatory Alliance (IVOA) defines standards and protocols that enable astronomers  to compare and cross-correlate these data sets in a consistent manner. A number of VO compatible tools already exist to find, query, manipulate such data. Tools also exist to process VO data via scripting languages (e.g., \textsc{voclient}). 

\begin{figure}
\begin{center}
\includegraphics[width=7cm]{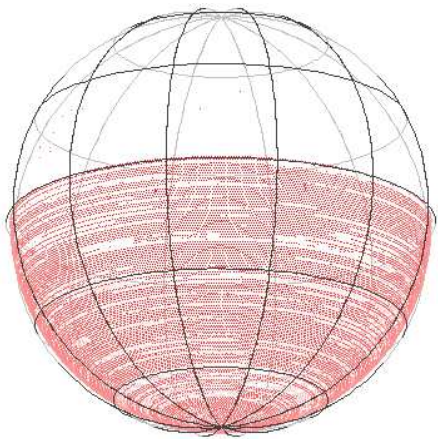}
\end{center}
\caption{Example screenshot from using the virtual observatory package TOPCAT.  This shows the positions (on the celestial sphere) of all observations for the 70\,cm pulsar survey.}\label{fg:vo}
\end{figure}

It is possible to query the metadata that provides information about each pulsar observation using VO tools.  Both cone-searches (allowing searches in position) and queries in the Astronomical Data Query Language  (ADQL) are implemented.  An example use-case would be to obtain a listing (in HTML, CSV or the more flexible VOTable format) of all files in the archive that were obtained in survey mode\footnote{ADQL is based upon a subset of SQL92 with extensions for astronomical usage.}.  The resulting VOTable can be loaded into virtual observatory packages (such as \textsc{TOPCAT}; Taylor 2005).  Figure~\ref{fg:vo} shows a TOPCAT display of the coordinates for all the observations in the 70\,cm pulsar survey.  A ``multi cone search" can then be run to match these search mode observations with, e.g., known pulsar positions from the ATNF pulsar catalogue (Manchester et al. 2005), or e.g., the AGILE catalogue of gamma-ray sources (Pittori et al. 2009)\footnote{Such a search can be carried out in TOPCAT by loading the resulting VO table from the ADQL query and then carrying out a multiple cone search with any of the catalogues that are currently in VO format.}.  One obvious possibility would be to select all pulsars with a specific property of interest from the ATNF pulsar catalogue (such as pulsars with high magnetic field strengths) and then use the virtual observatory tools to identify observations available for download that may help to study this class of pulsar.

\subsubsection{Large data sets}

The current data archive stores $\sim$5\,TB of data. The amount of data stored will increase rapidly as the data from more observing programmes are added.  It is clearly not possible to download a significant part of this archive using the online portals (currently a restriction of 50 files is placed on any individual download).   We are planning new approaches to allow access to such large data files using high performance computing infrastructure, but this has not yet been implemented.  Instead, for folded data sets the user may wish to obtain pre-processed files, which will avoid long download times.  The CSIRO data access portal provides the option to download the original or the pre-processed files.

\section{Using the data}

As each data file is stored in PSRFITS format, much of the standard software for processing FITS files can be used. For instance, the archiving software itself uses the \textsc{nom.tam} java library for reading the files\footnote{\url{http://heasarc.gsfc.nasa.gov/docs/heasarc/fits/java/v0.9/javadoc/}}. The NASA High Energy Astrophysics Science Archive Research Centre\footnote{\url{http://heasarc.gsfc.nasa.gov/docs/heasarc/fits.html}} provides many other tools that can be used. Available utility programs that work with PSRFITS include,
\begin{itemize}
\item \textsc{listhead} - This utility provides a listing of header parameters within the file.
\item \textsc{fitscopy} - Provides routines to copy FITS files (note that most options are not relevant for pulsar data)
\item \textsc{liststruc} - Lists the formatting internal to the FITS file (provides details on which parameters are stored as strings, integers, floating point, etc.)
\item \textsc{modhead} - Displays or modifies a header keyword.  For instance, this can be used to change the pulsar's name that is stored in the file.  For fold-mode files, the PSRCHIVE tool \textsc{psredit} can also be used for this purpose.
\item \textsc{tablist} - displays the contents of a FITS table.  This utility can be used to display tabular information from the FITS file; for instance, to determine the parallactic angle for each integration.
\item \textsc{tabcalc} - allows simple calculations to be performed on tables within the FITS file.  Columns may be overwritten or new columns created.  A new FITS file is created.
\item \textsc{fv} provides a graphical interface allowing the various header parameters and tables to be inspected by eye (and, if required, modified).  \textsc{fv} also provides simple plotting routines.  This is part of the much larger \textsc{FTOOLS} package which can be downloaded in its entirety. 
\end{itemize}
In general, only tools that work with general FITS data files are compatible with PSRFITS.  Utility programs that work with FITS images, e.g., \textsc{SAOIMAGE}, \textsc{DS9}, \textsc{IMLIST}, will not be compatible.

All fold-mode files can be processed using the PSR\-CHIVE  software suite.   A common sequence of processing steps would be to 1) download the data file using the CSIRO data access portal, 2) use \textsc{paz} and/or \textsc{pazi} to remove RFI, 3) \textsc{pac} to calibrate the profile, 4) \textsc{pam} to produce a single pulse profile integrated in observing frequency and over all integrations, 5) \textsc{pav} to view the pulse profile and 5) \textsc{pat} to obtain pulse times-of-arrival which can be processed using \textsc{tempo2}.

Search mode files can be processed using the \textsc{dspsr} (van Straten \& Bailes 2010) or \textsc{sigproc}\footnote{\url{http://sigproc.sourceforge.net/}. Note that only the most recent version of \textsc{sigproc} is compatible with PSRFITS.   It is expected that the next version of the \textsc{presto} search-mode package will also be compatible with our data files.} software packages. \textsc{sigproc} provides various tools for plotting the data or for searching for new pulsars.   \textsc{dspsr} allows the raw data to be displayed (using \textsc{searchplot}) or to be folded with a given period to form a folded profile (using \textsc{dspsr}).  

\subsection{Ancillary files}

The data archive provides access only to the observation data files.  In order to process these files it may be necessary to obtain extra data files relevant to the Parkes observatory.  For instance, the pulsar timing method requires that the clock used at the observatory to measure the pulse arrival times be converted to a realisation of terrestrial time.  This conversion is provided in a set of ``clock correction files'' that can be obtained as part of the \textsc{tempo2} distribution or from the pulsar web site\footnote{\url{http://www.atnf.csiro.au/research/pulsar}}.	Other useful files, such as measurements of the time delays between different backend instrumentation, may also be obtained from this website.

\subsection{Referencing the database}

Much of the data available from the archive is from on-going projects. Even though all data older than 18 months is out of any embargo period we recommend that the people who carried out the observations are contacted before extensive use is made of the data as each data set has its own peculiarities that may need to be understood.

Any publication containing these data sets should refer to the original paper describing the data sets.   We would also appreciate a reference to the portal used to download the data and/or a reference to this paper.  It is a requirement of the Australia Telescope National Facility that any publication making use of the Parkes data includes a specific acknowledgement that is listed on the CSIRO Astronomy and Space Science webpage\footnote{\url{http://www.atnf.csiro.au/research/publications}}.

\section{The future}

The initial data archive provides observations obtained from five observing programmes.  More than 300 different observing programmes relating to pulsars have been undertaken at the Parkes observatory and pulsar observations currently take up two-thirds of the total time on the telescope.   Work is on-going to ensure that all future observations are included in the archive. Owing to the volume of data it is unlikely that, in the near future, we will provide the data from an on-going Parkes pulsar survey (Keith et al. 2010).  When completed, this survey will require more than 1PB of data storage.  We are currently attempting to identify the means by which such large data sets could be stored, accessed and processed.  

After the software has been developed to include current observations in the archive, we will recover as many existing data sets as possible. The choice of which new observations to add into the archive depends upon data storage requirements and the accessibility of the data. It is likely that the next major data sets to be added will be 1) the Parkes multibeam survey, which discovered about half of all the known pulsars (Manchester et al. 2001), 2) the timing observations relating to new discoveries from this survey (Lorimer et al. 2006, Faulkner et al. 2004, Hobbs et al. 2004, Kramer et al. 2003, Morris et al. 2002, Manchester et al. 2001) and 3) the timing observations being carried out as part of the Fermi gamma-ray mission (Weltevrede et al. 2010).   A list of the data sets currently available is on our website\footnote{\url{http://www.atnf.csiro.au/research/pulsar/index.php?n=Main.ANDSATNF}}.  

In the near future, it is likely that observations from a 12-m antenna commissioned in 2008 at the Parkes Observatory as a test-bed for new technology receivers for the Australian Square Kilometre Array Pathfinder (ASKAP) will be included as part of the archive.  In the longer term it is possible that our data archive will merge with the Australia Telescope Online Archive\footnote{\url{http://atoa.atnf.csiro.au/}} and provide observations from Parkes, the Australia Telescope Compact Array and the Mopra telescopes.

\section{Conclusions}

Observations at the Parkes radio telescope have led to numerous discoveries relating to pulsar astrophysics.  The data archive described here allows, for the first time, access to many of the original observations that were used in making these discoveries. It is hoped that this new resource will be used for numerous scientific projects including long-term pulsar timing experiments, discovering new pulsars in existing data sets and to provide an archive of high time-resolution data allowing new and unexpected discoveries.

\section*{Acknowledgments} 

This project is supported by the Australian National Data Service (ANDS). ANDS is supported by the Australian Government through the National Collaborative Research Infrastructure Strategy Program and the Education Investment Fund (EIF) Super Science Initiative (\url{http://www.ands.org.au}).  We acknowledge the software development provided by the CSIRO IM\&T Software Services, the business process development by the CSIRO IM\&T Data Management Service and project management through Citadel Systems. This research has made use of software provided by the UK's AstroGrid Virtual Observatory Project, which is funded by the Science and Technology Facilities Council and through the EU's Framework 6 programme. The data archive relies on data that have been obtained and processed by numerous people.  In particular we acknowledge the work undertaken by A. Teoh, M. Hobbs, R. Neil and D. Smith. The Parkes radio telescope is part of the Australia Telescope, which is funded by the Commonwealth of Australia for operation as a National Facility managed by the Commonwealth Scientific and Industrial Research Organisation (CSIRO).  GH is the recipient of an Australian Research Council QEII Fellowship (\#DP0878388).






\end{document}